\documentclass[fleqn,twoside,11pt]{article}

\usepackage{amssymb}
\usepackage{graphicx}
\usepackage{amsmath}

\begin{document}

\noindent
{\sf University of Shizuoka}

\hspace*{13cm} {\large US-07-01}

\vspace{3mm}

\begin{center}

{\Large\bf  A$_4$ Symmetry and Lepton Masses and Mixing}

\vspace{3mm}
{\bf Yoshio Koide}

{\it Department of Physics, University of Shizuoka, 
52-1 Yada, Shizuoka 422-8526, Japan\\
E-mail address: koide@u-shizuoka-ken.ac.jp}

\date{\today}
\end{center}

\begin{abstract}
Stimulated by Ma's idea which explains the tribimaximal 
neutrino mixing by assuming an A$_4$ flavor symmetry,
a lepton mass matrix model is investigated.
A Frogatt-Nielsen type model is assumed, and the
flavor structures of the masses and mixing are caused by the VEVs of
SU(2)$_L$-singlet scalars $\phi_i^u$ and $\phi_i^d$ ($i=1,2,3$),
which are assigned to {\bf 3} and $({\bf 1}, {\bf 1}',{\bf 1}'')$
of A$_4$, respectively.
\end{abstract}

\vspace{3mm}


{\large\bf 1 \ Introduction}

It is generally considered that masses and mixings of the quarks and leptons 
will obey a simple law of nature. Then, it is also likely that the masses and
mixings of those fundamental particles will be governed by a symmetry. 
However, even if there is such a simple relation in the quark 
sector, it is hard to see such a relation in the quark sector, 
because the original symmetry will be spoiled by the gluon cloud. 
Therefore, in the present paper, we will confine ourselves 
to the investigation of the lepton masses and mixings.

It is well known that the observed neutrino mixing is 
nearly described by the so called tribimaximal mixing \cite{tribi}
$$
U_{TB}=\left(
\begin{array}{ccc}
\frac{2}{\sqrt6} & \frac{1}{\sqrt3} & 0 \\
-\frac{1}{\sqrt6} & \frac{1}{\sqrt3} & -\frac{1}{\sqrt2} \\
-\frac{1}{\sqrt6} & \frac{1}{\sqrt3} & \frac{1}{\sqrt2}
\end{array} \right) .
\eqno(1.1)
$$
In order to understand the tribimaximal mixing,  Ma \cite{Ma06} 
has, recently, 
proposed a neutrino mass matrix model based on a non-Abelian 
discrete symmetry A$_4$. 
The symmetry A$_4$ seems to be very promising for a model of the leptons.

On the other hand, it is also well known that the observed charged lepton 
masses satisfy the relation \cite{Koidemass,Koide90}
$$
m_e+m_\mu+m_\tau=\frac{2}{3}(\sqrt{m_e}+\sqrt{m_\mu}+\sqrt{m_\tau})^2 ,
\eqno(1.2)
$$
with remarkable precision. 
The mass formula (1.2) is invariant under any 
exchange $\sqrt{m_i}\leftrightarrow \sqrt{m_j}$ ($i,j=e,\mu,\tau$). 
This 
suggests that the lepton mass matrix model will be described 
by a permutation symmetry S$_3$ \cite{S3}.

In order to understand the formula (1.2), a seesaw-type mass matrix model 
\cite{Koide90,KF96,KT96} has been proposed:
$$
M_e=m_L^e M_E^{-1}m_R^e .
\eqno(1.3)
$$
Here, $M_E$ is a mass matrix of hypothetical heavy leptons $E_i$ $(i=1,2,3)$, 
and we have assumed $M_E \propto {\bf 1}\equiv {\rm diag}(1,1,1)$.
The matrices $m_L^e$ and $m_R^e$ are mass matrices defined by 
$\overline{e}_L m_L^e E_R$ and $\overline{E}_L m_R^e e_R$, respectively, 
and we assume $m_L^e=m_R^e/k =y_e {\rm diag}(v_1, v_2, v_3)$ 
($k$ is a constant), where $v_i$ are vacuum 
expectation values (VEVs) of 3 Higgs scalars $\phi_{Li}=(\phi_{Li}^\dagger, 
\phi_{Li}^0)$, and they satisfy the relation
$$
v_1^2+v_2^2+v_3^2=\frac{2}{3}(v_1+v_2+v_3)^2 .
\eqno(1.4)
$$
The relation (1.4) can be derived from the following Higgs potential 
\cite{Koide99,Koide06}
$$
V=\mu^2(\phi_1^\dagger \phi_1+\phi_2^\dagger \phi_2+\phi_3^\dagger \phi_3)
+\frac{1}{2}\lambda_1(\phi_1^\dagger \phi_1+\phi_2^\dagger \phi_2
+\phi_3^\dagger \phi_3)^2
$$
$$
+\lambda_2(\phi_\sigma^\dagger \phi_\sigma)(\phi_\pi^\dagger \phi_\pi
+\phi_\eta^\dagger \phi_\eta)+V_{SB} ,
\eqno(1.5)
$$
where $\phi_i$ $(i=1,2,3)$ are 3 objects of S$_3$
(fundamental basis),  $(\phi_\pi,\phi_\eta)$ 
and $\phi_\sigma$ are doublet and singlet of S$_3$, respectively, which are 
defined by
$$
\left(\begin{array}{c}
\phi_\pi \\
\phi_\eta \\
\phi_\sigma
\end{array} \right)=
\left(\begin{array}{ccc}
0 & -\frac{1}{\sqrt2} & \frac{1}{\sqrt2} \\
\frac{2}{\sqrt6} & -\frac{1}{\sqrt6} & -\frac{1}{\sqrt6} \\
\frac{1}{\sqrt3} & \frac{1}{\sqrt3} & \frac{1}{\sqrt3}
\end{array} \right)
\left(\begin{array}{c}
\phi_1 \\
\phi_2 \\
\phi_3
\end{array} \right) ,
\eqno(1.6)
$$
and $V_{SB}$ is a soft symmetry breaking term \cite{Koide06}
which does not affect the derivation of the relation (1.4).
The minimizing condition of the potential (1.5) leads to the VEV relation
$$
v_\pi^2+v_\eta^2=v_\sigma^2 .
\eqno(1.7)
$$
The relation (1.7) gives the relation (1.4) because 
$$
v_1^2+v_2^2+v_3^2=v_\pi^2+v_\eta^2+v_\sigma^2=2v_\sigma^2
 = 2 \left( \frac{v_1+v_2+v_3}{\sqrt{3}}\right)^2 .
\eqno(1.8)
$$
(Note that although the Higgs potential (1.5) is invariant
under the S$_3$ symmetry, but it is not a general one of the 
S$_3$ invariant form. 
As pointed out in Ref.~\cite{Koide06}, a Higgs potential with
the general form cannot lead to the relation (1.7).
We need an additional requirement.)
For a recent S$_3$ model of the lepton masses and mixings, see 
Ref.~\cite{Koide0605}.

Considering the scenario for the charged 
lepton mass spectrum, the S$_3$ symmetry is also attractive, but,
in the present paper, we will investigate an A$_4$ model by noticing 
Ma's model \cite{Ma06} for the tribimaximal neutrino mixing.
In the next section, we will show that the S$_3$ scenario for the 
charged lepton masses can be translated into the language of A$_4$.
In Sec.3, we will give a Frogatt-Nielsen \cite{Frogatt} type 
model of the leptons based on an A$_4$ flavor symmetry.
The mass matrix structures in the charged lepton and neutrino 
sectors are discussed in Secs.4 and 5, respectively.
In Sec.6, a speculation about the neutrino masses will be given. 
In Sec.7, a SUSY version of the Higgs potential (1.5) [(2.16)] will
be proposed.
Finally, Sec.8 is devoted to the summary.
In order to obtain the tribimaximal mixing (1.1) and the charged
lepton mass relation (1.2), we will need further phenomenological
assumptions, (i) ${\bf 1}' \leftrightarrow {\bf 1}''$ symmetry and
(ii) the universality of the coupling constants, in addition
to the A$_4$ symmetry. 

\vspace{3mm}

{\large\bf 2 \ From S$_3$ into A$_4$}

When we define $\overline{\psi}=(\overline{\psi}_1, \overline{\psi}_2, 
\overline{\psi}_3)$ and $\psi=(\psi_1, \psi_2, \psi_3)$ as {\bf 3} of 
A$_4$, we can compose ${\bf 1}$, ${\bf 1}'$ and ${\bf 1}''$ of A$_4$ 
as follows:
$$
(\overline{\psi}\psi)_{\bf 1}=\frac{1}{\sqrt3}(\overline{\psi}_1 \psi_1
+\overline{\psi}_2 \psi_2+\overline{\psi}_3 \psi_3) ,
\eqno(2.1)
$$
$$
(\overline{\psi}\psi)_{{\bf 1}'}=\frac{1}{\sqrt3}(\overline{\psi}_1 \psi_1
+\overline{\psi}_2 \psi_2 \omega+\overline{\psi}_3 \psi_3 \omega^2) ,
\eqno(2.2)
$$
$$
(\overline{\psi}\psi)_{{\bf 1}''}=\frac{1}{\sqrt3}(\overline{\psi}_1 \psi_1
+\overline{\psi}_2 \psi_2 \omega^2+\overline{\psi}_3 \psi_3 \omega) ,
\eqno(2.3)
$$
where
$$
\omega=e^{i \frac{2}{3}\pi}=\frac{-1+i\sqrt{3}}{2} .
\eqno(2.4)
$$
The expressions (2.1)--(2.3) can be rewritten as 
$$
(\overline{\psi}\psi)_{\bf 1}=(\overline{\psi}\psi)_\sigma ,
\eqno(2.5)
$$
$$
(\overline{\psi}\psi)_{{\bf 1}'}=\frac{1}{\sqrt2}[(\overline{\psi}\psi)_\eta-i
(\overline{\psi}\psi)_\pi] ,
\eqno(2.6)
$$
$$
(\overline{\psi}\psi)_{{\bf 1}''}=\frac{1}{\sqrt2}[(\overline{\psi}\psi)_\eta+i
(\overline{\psi}\psi)_\pi] ,
\eqno(2.7)
$$
where
$$
\left(\begin{array}{c}
(\overline{\psi}\psi)_\sigma \\
(\overline{\psi}\psi)_\eta \\
(\overline{\psi}\psi)_\pi
\end{array} \right)=
\left(\begin{array}{ccc}
\frac{1}{\sqrt3} & \frac{1}{\sqrt3} & \frac{1}{\sqrt3} \\
\frac{2}{\sqrt6} & -\frac{1}{\sqrt6} & -\frac{1}{\sqrt6} \\
0 & -\frac{1}{\sqrt2} & \frac{1}{\sqrt2}
\end{array} \right)
\left(\begin{array}{c}
\overline{\psi}_1 \psi_1 \\
\overline{\psi}_2 \psi_2 \\
\overline{\psi}_3 \psi_3
\end{array} \right) .
\eqno(2.8)
$$
It is useful to define the following $(\phi_\sigma, \phi_\eta, \phi_\pi)$
basis correspondingly to (2.5)--(2.7), 
$$
\phi_{\bf 1}=\phi_\sigma ,
\eqno(2.9)
$$
$$
\phi_{{\bf 1}'}=\frac{1}{\sqrt2}(\phi_\eta-i \phi_\pi) ,
\eqno(2.10)
$$
$$
\phi_{{\bf 1}''}=\frac{1}{\sqrt2}(\phi_\eta+i \phi_\pi) ,
\eqno(2.11)
$$
where the scalars $\phi_{\bf 1}$, $\phi_{{\bf 1}'}$ and $\phi_{{\bf 1}''}$ 
are ${\bf 1}$, ${\bf 1}'$ and ${\bf 1}''$ of A$_4$.
Then, A$_4$-invariant Yukawa interactions which are composed of 
$\overline{\psi}, \psi$ and $\phi$ are expressed as follows:
$$
(\overline{\psi}\psi)_{\bf 1}\phi_{\bf 1}=(\overline{\psi}\psi)_\sigma 
\phi_\sigma ,
\eqno(2.12)
$$
$$
(\overline{\psi}\psi)_{{\bf 1}'} \phi_{{\bf 1}''}=\frac{1}{2}[(\overline{\psi}
\psi)_\eta \phi_\eta+(\overline{\psi}\psi)_\pi \phi_\pi+i(\overline{\psi}\psi)
_\eta \phi_\pi-i(\overline{\psi}\psi)_\pi \phi_\eta] ,
\eqno(2.13)
$$
$$
(\overline{\psi}\psi)_{{\bf 1}''} \phi_{{\bf 1}'}=\frac{1}{2}[(\overline{\psi}
\psi)_\eta \phi_\eta+(\overline{\psi}\psi)_\pi \phi_\pi
-i(\overline{\psi}\psi)_\eta \phi_\pi+i(\overline{\psi}\psi)_\pi \phi_\eta] .
\eqno(2.14)
$$
Hereafter, we will always assume a ${\bf 1}' \leftrightarrow {\bf 1}''$ 
symmetry, so that we obtain
$$
(\overline{\psi}\psi)_{{\bf 1}'} \phi_{{\bf 1}''}+(\overline{\psi}\psi)
_{{\bf 1}''} \phi_{{\bf1}'}=(\overline{\psi}\psi)_\eta \phi_\eta+(\overline
{\psi}\psi)_\pi \phi_\pi .
\eqno(2.15)
$$

In the S$_3$-invariant Higgs potential (1.5), the existence of the 
$\lambda_2$-term was essential for the derivation of the VEV relation (1.7). 
In the present A$_4$ model, if we adopt the basis 
$\phi=(\phi_\sigma, \phi_\eta, \phi_\pi)$ which are defined by 
Eqs.(2.9)--(2.11), we can regard the Higgs 
potential (1.5) as an A$_4$-invariant one:
$$
V=\mu^2(\phi_{\bf 1}^\dagger \phi_{\bf 1}
+\phi_{{\bf 1}'}^\dagger \phi_{{\bf 1}''}
+\phi_{{\bf 1}''}^\dagger \phi_{{\bf 1}'})
+\frac{1}{2}\lambda_1 (\phi_{\bf 1}^\dagger \phi_{\bf 1}
+\phi_{{\bf 1}'}^\dagger \phi_{{\bf 1}''}
+\phi_{{\bf 1}''}^\dagger \phi_{{\bf 1}'})^2
$$
$$
+\lambda_2 (\phi_{\bf 1}^\dagger \phi_{\bf 1})
(\phi_{{\bf 1}'}^\dagger \phi_{{\bf 1}''}
+\phi_{{\bf 1}''}^\dagger \phi_{{\bf 1}'})
+V_{SB} 
$$
$$
 =\mu^2(\phi_\sigma^\dagger \phi_\sigma+\phi_\eta^\dagger \phi_\eta
+\phi_\pi^\dagger \phi_\pi)
+\frac{1}{2}\lambda_1 (\phi_\sigma^\dagger \phi_\sigma
+\phi_\eta^\dagger \phi_\eta+\phi_\pi^\dagger \phi_\pi)^2
$$
$$
+\lambda_2(\phi_\sigma^\dagger \phi_\sigma)(\phi_\pi^\dagger \phi_\pi
+\phi_\eta^\dagger \phi_\eta)+V_{SB} .
\eqno(2.16)
$$
When we define 
$(\phi_1, \phi_2, \phi_3)$ by Eq.(1.6), we can obtain the VEV relation (1.4) 
for the VEVs $v_i=\langle \phi_i \rangle$, so that we obtain the charged 
lepton mass relation (1.2) from the A$_4$-invariant Yukawa interaction
$$
(\overline{e}E)_{\bf 1} \phi_{\bf 1} 
+(\overline{e}E)_{{\bf 1}'} \phi_{{\bf 1}''}
+(\overline{e}E)_{{\bf 1}''} \phi_{{\bf 1}'} =
(\overline{e}E)_\sigma \phi_\sigma +(\overline{e}E)_\eta \phi_\eta
+(\overline{e}E)_\pi \phi_\pi
$$
$$
=\overline{e}_1 E_1 \phi_1+\overline{e}_2 E_2 \phi_2+\overline{e}_3 
E_3 \phi_3 ,
\eqno(2.17)
$$
where $e_L$ and $E_R$ have been assigned to {\bf 3} of A$_4$.
(However, in the next section, we will not adopt the seesaw model
(1.3), but do a Frogatt-Nielsen type model without the heavy leptons $E_i$.)

Note that, in the S$_3$ model, $(\phi_1, \phi_2, \phi_3)$ 
were 3 objects of S$_3$ and $(\phi_\sigma, \phi_\eta, \phi_\pi)$ were 
(singlet, doublet) of S$_3$, while, in the present A$_4$ model, $(\phi_\sigma, 
\phi_\eta, \phi_\pi)$ and $(\phi_1, \phi_2, \phi_3)$ are merely linear 
combinations of $(\phi_{\bf 1}, \phi_{{\bf 1}'}, \phi_{{\bf 1}''})$, and they 
are not irreducible representations of A$_4$.

Thus, we have a possibility that we can build a model which leads not only to 
the tribimaximal mixing for the neutrinos, but also to the mass relation 
(1.2) for the charged leptons by developing Ma's idea.

\vspace{3mm}

{\large\bf 3 \ Model}

So far, we have considered that 3 scalars $\phi_i$ are SU(2) doublets. 
However, such a model with multi-Higgs doublets causes a flavor changing 
neutral current (FCNC) problem. Therefore, in the present paper, we assume a 
Frogatt-Nielsen \cite{Frogatt} type model
$$
H_{eff}=y_e \overline{l}_L H_L^d \frac{\phi^d}{\Lambda_d}
\frac{\phi^d}{\Lambda_d}e_R
+y_\nu \overline{l}_L H_L^u \frac{\phi^u}{\Lambda_u} \nu_R+y_R 
\overline{\nu}_R 
\Phi\nu_R^\ast ,
\eqno(3.1)
$$
where $\ell_{iL}$ are SU(2)$_L$-doublet leptons 
$\ell_{iL}=(\nu_{iL}, e_{iL})$, 
$H_L^d$ and $H_L^u$ are conventional SU(2)$_L$ doublet Higgs scalars, 
$\phi^d$ and $\phi^u$ are SU(2)$_L$ singlet scalars, and $\Lambda_d$ and 
$\Lambda_u$ are scales of the effective theory. 
We consider that $\langle \phi^f \rangle/\Lambda_f$ ($f=u,d$) are
of the order of 1. 
Here, we have not adopted the seesaw 
type model (1.3) for the charged lepton sector, because the existence of $M_E 
\propto {\bf 1}$ in Eq.(1.3) did not play any essential role in the flavor 
structure of the charged lepton mass matrix $M_e$. The scalar $\Phi$ has 
been introduced in order to generate the Majorana mass $M_R$ of the 
right-handed neutrinos $\nu_R$.
The model is essentially unchanged compared with the seesaw model
as far as the flavor structures are concerned.
However, the scenario for the energy scale of the symmetry breaking 
is considerably changed, i.e. we consider that the VEVs of $\phi_i^f$ 
are of the order of the Planck mass scale although we have considered 
$\langle \phi_i \rangle \sim 10^2$ GeV in the seesaw model 
\cite{KoideEPJC06}. 
In other words, in the Frogatt-Nielsen-type model, the A$_4$-broken 
structure of the effective Yukawa coupling constants is formed
 at the Planck mass scale.
However, this is not serious problem, because the formula
(1.1) is not so sensitive to the renormalization group equation (RGE) 
effects as far as  the lepton sector is concerned \cite{evol_KoideF}.
Although the relation (1.2) is in remarkable agreement with the
observed charged lepton masses (the pole masses), the standpoint in
the present paper is that the remarkable coincidence is accidental
and the relation (1.2) will be satisfied only approximately at a
low energy scale.

Ma  has assigned the scalars $\phi^d$ to {\bf 3} of A$_4$
in Ref.\cite{Ma06}. 
However, as we have shown in Sec.2, since the 
scalars which can give the VEV relation (1.7) [or (1.4)] are not {\bf 3} of 
A$_4$, but $({\bf 1},{\bf 1}',{\bf 1}'')$ of A$_4$, we regard $\phi^d$ as 
$({\bf 1},{\bf 1}',{\bf 1}'')$ of A$_4$ in the present paper. 
Also, we will change the assignment of $e_R$ and $\nu_R$ from those
in the Ma model. 
Of course, the essential idea to 
obtain the tribimaximal mixing is indebted to the Ma model. 
The A$_4$ assignments in the present model are listed in Table 1.
In order to 
forbid unwelcome combinations $\overline{l}_L H_L^d(\phi^d)^n(\phi^u)^m e_R$ 
except for $(n,m)=(2,0)$ and $\overline{l}_L H_L^u(\phi^d)^n(\phi^u)^m \nu_R$ 
except for $(n,m)=(0,1)$, for example, we assume the following U(1) charge 
assignments: $Q(l_L)=Q(\nu_R)=Q(e_R)=Q(\Phi)=0$, 
$Q(\phi^d)=-\frac{1}{2}Q(H_l^d)=q_d >0$, and $Q(\phi^u)=-Q(H_L^u)=q_u >0$, 
where $q_d/q_u \neq n/2$ and $q_u/q_d \neq n$ $(n=0,1,2,3,\cdots)$.

\vspace{3mm}

{\large\bf 4 \ Charged lepton sector}

In the charged lepton sector, the possible A$_4$-invariant interactions 
$(\overline{e}_L e_R)\phi^d \phi^d$, i.e., $({\bf 3}\times{\bf 3})
\times({\bf 1},{\bf 1}',{\bf 1}'')\times({\bf 1},{\bf 1}',{\bf 1}'')$, are 
given by
$$
H_e =(\overline{e}e)_{\bf 1}[y_0\phi_{\bf 1}\phi_{\bf 1}+y_1
(\phi_{\bf 1}'\phi_{\bf 1}''+\phi_{\bf 1}''\phi_{\bf 1}')]
+y_2[(\overline{e}e)_{\bf 1}'\phi_{\bf 1}'\phi_{\bf 1}'
+(\overline{e}e)_{\bf 1}''\phi_{\bf 1}''\phi_{\bf 1}'']
$$
$$
+y_3[(\overline{e}e)_{\bf 1}'(\phi_{\bf 1}''\phi_{\bf 1}+\phi_{\bf 1}
\phi_{\bf 1}'')+(\overline{e}e)_{\bf 1}''(\phi_{\bf 1}'\phi_{\bf 1}
+\phi_{\bf 1}\phi_{\bf 1}')]
$$
$$
=(\overline{e}e)_\sigma[y_0\phi_\sigma^2+y_1(\phi_\pi^2+\phi_\eta^2)]
+\frac{1}{\sqrt2}y_2[(\overline{e}e)_\eta(\phi_\eta^2-\phi_\pi^2)
-2(\overline{e}e)_\pi\phi_\eta\phi_\pi]
$$
$$
+2y_3[(\overline{e}e)_\eta\phi_\eta\phi_\sigma+(\overline{e}e)_\pi\phi_\pi
\phi_\sigma] ,
\eqno(4.1)
$$
where, for convenience, we have dropped the index $d$, and we have assumed the 
${\bf 1}' \leftrightarrow {\bf 1}''$ symmetry. Furthermore, if we assume the 
universality of the coupling constants,
$$
y_0=y_1=y_2=y_3 ,
\eqno(4.2)
$$
we obtain 
$$
H_e=y_0 \left[(\overline{e}e)_\sigma(\phi_\sigma^2+\phi_\pi^2+\phi_\eta^2)
+\frac{1}{\sqrt2}(\overline{e}e)_\eta(\phi_\eta^2-\phi_\pi^2
+2\sqrt2\phi_\eta\phi_\sigma)
+(\overline{e}e)_\pi(-2\phi_\eta\phi_\pi+2\sqrt2\phi_\pi\phi_\sigma) \right] 
$$
$$
=y_0 \left[(\overline{e}e)_\sigma(\phi_1^2+\phi_2^2+\phi_3^2)
+\frac{1}{\sqrt2}(\overline{e}e)_\eta(2\phi_1^2-\phi_2^2-\phi_3^2)
+\sqrt{\frac{3}{2}}(\overline{e}e)_\pi(\phi_3^2-\phi_2^2) \right]
$$
$$
=\sqrt3y_0(\overline{e}_1 e_1 \phi_1^2+\overline{e}_2 e_2 \phi_2^2
+\overline{e}_3 e_3 \phi_3^2) ,
\eqno(4.3)
$$
where $\phi_i$ $(i=1,2,3)$ are defined by Eq.(1.6). As we discussed in Sec.2, 
since we can write the Higgs potential (1.5) for $\phi^d=(\phi_\pi^d, 
\phi_\eta^d, \phi_\sigma^d)$, we can obtain the VEV relation (1.7) [i.e. (1.4) 
for $v_i=\langle \phi_i^d \rangle$]. 
Therefore, from Eqs.(4.3) and (1.4), we 
can obtain the charged lepton mass relation (1.2).

\vspace{3mm}

{\large\bf 5 \ Neutrino sector}

Since $\overline{\nu}_L \phi^u \nu_R \sim {\bf 3}\times{\bf 3}\times({\bf 1},
 {\bf 1}', {\bf 1}'')$, the A$_4$-invariant Yukawa interactions 
are as follows:
$$
H_\nu =y_0^\nu(\overline{\nu}_L \phi^u)_\sigma \nu_{R\sigma}
+y_1^\nu \left[(\overline{\nu}_L \phi^u)_\eta \nu_{R\eta}
+(\overline{\nu}_L \phi^u)_\pi \nu_{R\pi} \right] ,
\eqno(5.1)
$$
so that we obtain the mass matrix $m_L^\nu$ which is defined by 
$$
(\overline{\nu}_{1} \ \ \overline{\nu}_{2} \ \ \overline{\nu}_{3})_L
m_L^\nu
\left( \begin{array}{c}
\nu_{\sigma} \\
\nu_{\eta} \\
\nu_{\pi}
\end{array} \right)_R ,
\eqno(5.2)
$$
as follows:
$$
m_L^\nu=\left( \begin{array}{ccc}
\frac{1}{\sqrt3}y_0^\nu v_i^u & y_1^\nu \frac{2}{\sqrt6}v_1^u & 0 \\
\frac{1}{\sqrt3}y_0^\nu v_2^u & -y_1^\nu \frac{1}{\sqrt6}v_2^u 
& -y_1^\nu \frac{1}{\sqrt2}v_2^u \\
\frac{1}{\sqrt3}y_0^\nu v_3^u & -y_1^\nu \frac{1}{\sqrt6}v_3^u 
& y_1^\nu \frac{1}{\sqrt2}v_3^u
\end{array} \right) .
\eqno(5.3)
$$
When we again assume the universality of the coupling constants, 
$y_0^\nu=y_1^\nu$, we obtain 
$$
m_L^\nu=DU_{TB} ,
\eqno(5.4)
$$
$$
D=y_0^\nu {\rm diag}(v_1^u, v_2^u, v_3^u) ,
\eqno(5.5)
$$
and $U_{TB}$ is the tribimaximal mixing matrix (1.1), where we have changed 
the basis of $\nu_R$ from $(\nu_\sigma, \nu_\eta, \nu_\pi)_R$ to 
$(\nu_\eta, \nu_\sigma, \nu_\pi)_R$.

Although the VEVs of the scalars $\phi^d=(\phi_\pi^d, \phi_\eta^d, 
\phi_\sigma^d)$ satisfy the VEV relation (1.7), 
the VEVs $v_i^u=\langle \phi_i^u \rangle$ do not have such a relation, 
because we cannot write an 
A$_4$-invariant term which corresponds to the $\lambda_2$-term, i.e. 
$\phi_\sigma^2(\phi_\pi^2+\phi_\eta^2)$. The potential for the scalars 
$\phi^u=(\phi_1^u, \phi_2^u, \phi_3^u)$ is symmetric for any exchange 
$\phi_i^u \leftrightarrow \phi_j^u$. Therefore, we consider
$$
\langle \phi_1^u \rangle=\langle \phi_2^u \rangle=\langle \phi_3^u \rangle 
\equiv v_u .
\eqno(5.6)
$$
Then, the mass matrix $m_L^\nu$ is diagonalized as 
$$
U_{TB}^T m_L^\nu=y_0^\nu v_u {\bf 1} .
\eqno(5.7)
$$
If the Majorana mass matrix $M_R$ is diagonal on the basis $(\nu_{\eta}, 
\nu_{\sigma}, \nu_{\pi})_R$, i.e.
$$
M_R={\rm diag}(M_\eta, M_\sigma, M_\pi) ,
\eqno(5.8)
$$
we obtain the mixing matrix $U_{MNS}$ and the eigenvalues $m_{\nu i}$ of the 
neutrino mass matrix $M_\nu=m_L^\nu M_R^{-1}(m_L^\nu)^T$,
$$
U_{MNS}=U_{TB} ,
\eqno(5.9)
$$
$$
m_{\nu 1}=(y_0^\nu v_u)^2 \frac{1}{M_\eta} , \ \ m_{\nu 2}=(y_0^\nu v_u)^2 
\frac{1}{M_\sigma}, \ \ m_{\nu 3}=(y_0^\nu v_u)^2 \frac{1}{M_\pi} .
\eqno(5.10)
$$
The explicit structure of $M_R={\rm diag}(M_\eta, M_\sigma, M_\pi)$ will be 
discussed in the next section.

\vspace{3mm}

{\large\bf 6 \ Speculation on the neutrino mass spectrum}

In order to speculate on the neutrino mass spectrum, let us assume that the 
Majorana masses are generated by the following interaction with
the scalars $\Phi =(\Phi_{\bf 1}, \Phi_{{\bf 1}'}, \Phi_{{\bf 1}''})$,
$$
H_{R}= \left[ y_0^R \overline{\nu}_{\bf 1} \nu^*_{\bf 1}
+y_1^R(\overline{\nu}_{{\bf 1}'}\nu^*_{{\bf 1}''}
+\overline{\nu}_{{\bf 1}''} \nu^*_{{\bf 1}'}) \right] \Phi_{\bf 1}
+y_2^R \left(\overline{\nu}_{{\bf 1}'}\nu^*_{{\bf 1}'} \Phi_{{\bf 1}'}
+\overline{\nu}_{{\bf 1}''} \nu^*_{{\bf 1}''} \Phi_{{\bf 1}''} \right) 
$$
$$
=\left[y_0^R \overline{\nu}_\sigma \nu_\sigma^\ast
+y_1^R(\overline{\nu}_\pi \nu_\pi^\ast +\overline{\nu}_\eta \nu_\eta^\ast) 
\right] \Phi_\sigma
+\frac{1}{\sqrt2}y_2^R \left[(\overline{\nu}_\eta \nu_\eta^\ast
-\overline{\nu}_\pi \nu_\pi^\ast)\Phi_\eta
-(\overline{\nu}_\pi \nu_\eta^\ast+\overline{\nu}_\eta \nu_\pi^\ast)\Phi_\pi 
\right] .
\eqno(6.1)
$$
We assume that the VEVs of $\Phi \sim ({\bf 1}, {\bf 1}', {\bf 1}'')$ 
satisfy the same relation as (1.7) for $\phi^d$
$$
\langle \Phi_\pi \rangle^2+\langle \Phi_\eta \rangle^2=\langle \Phi_\sigma 
\rangle^2 ,
\eqno(6.2)
$$
because $\phi^d$ and $\Phi$ are assigned to the same multiplets 
$({\bf 1}, {\bf 1}', {\bf 1}'')$.
However, if we consider $\langle \Phi_\pi \rangle \neq 0$, the matrix
$M_R$ cannot become diagonal.
Therefore, we assume $\langle \Phi_\pi \rangle = 0$, so that we will take
$$
\langle \Phi_\pi \rangle = 0, \ \ \ 
\langle \Phi_\eta \rangle=\langle \Phi_\sigma \rangle .
\eqno(6.3)
$$
This assumption corresponds to that, although we have already assumed the 
${\bf 1}' \leftrightarrow {\bf 1}''$ symmetry, we have assumed this symmetry
for the VEV values of $\Phi$, not for the fields, i.e. 
$\langle \Phi_{{\bf 1}'}\rangle =\langle\Phi_{{\bf 1}''}\rangle 
=\langle \Phi_\eta \rangle/\sqrt{2}$.
Then, we obtain the eigenvalues of $M_R$ as follows:
$$
M_\eta=\left( y_1^R +\frac{1}{\sqrt2}y_2^R\right) \langle \Phi_\sigma \rangle,
 \ \ 
M_\sigma=y_0^R \langle \Phi_\sigma \rangle  , \ \ 
M_\pi=\left( y_1^R -\frac{1}{\sqrt2}y_2^R\right) \langle \Phi_\sigma \rangle .
\eqno(6.4)
$$

In order to speculate the neutrino masses $m_{\nu i}$, we must more reduce 
the number of the parameters.
Therefore, let us assume that the fermion terms 
which couple to the scalars $\Phi_\sigma$, $\Phi_\eta$ and $\Phi_\pi$ are 
normalized as
$$
H_R=y_R \left[\left(\sin \alpha\, \overline{\nu}_\sigma \nu_\sigma^\ast
+\cos\alpha\, \frac{\overline{\nu}_\pi \nu_\pi^\ast +\overline{\nu}_\eta 
\nu_\eta}{\sqrt2} \right)\Phi_\sigma+\frac{\overline{\nu}_\eta \nu_\eta^\ast
-\overline{\nu}_\pi \nu_\pi^\ast}{\sqrt2}\Phi_\eta-\frac{\overline{\nu}_\pi 
\nu_\eta^\ast+\overline{\nu}_\eta \nu_\pi^\ast}{\sqrt2}\Phi_\pi \right] .
\eqno(6.5)
$$
Then, we can write the eigenvalues (6.4) as follows:
$$
M_\eta=\frac{y_R}{\sqrt2}  \langle \Phi_\sigma \rangle (1+\cos\alpha) , \ \ 
M_\sigma=y_R \langle \Phi_\sigma \rangle  , \ \ 
|M_\pi|=\frac{y_R}{\sqrt2} \langle \Phi_\sigma \rangle (1-\cos\alpha) ,
\eqno(6.6)
$$
so that we obtain the neutrino mass spectrum
$$
m_{\nu1}=\frac{1}{1+\cos\alpha}m_{\nu0} , \ \ 
m_{\nu2}=\frac{1}{\sqrt2 \sin\alpha}m_{\nu0} , \ \ 
m_{\nu3}=\frac{1}{1-\cos\alpha}m_{\nu0} ,
\eqno(6.7)
$$
where $m_{\nu0}=(y_0^\nu v_u)^2 / (y_R \langle 
\Phi_\sigma \rangle / \sqrt2)^2$.
For the observed ratio \cite{solar,atm}
$$
R_{obs} \equiv \frac{\Delta m^2 _{solar}}{\Delta m^2_{atm}}
=\frac{(7.9^{+0.6}_{-0.5})\times 10^{-5}{\rm eV}^2}{(2.74^{+0.44}_{-0.26})
\times 10^{-3} {\rm eV}^2}
=(2.9\pm0.5)\times 10^{-2} ,
\eqno(6.8)
$$
we obtain the predicted ratio
$$
R=\frac{\Delta m^2_{21}}{\Delta m^3_{32}}
=\frac{m_{\nu 2}^2-m_{\nu 1}^2}{m_{\nu 3}^2-m_{\nu 2}^2}
=\frac{(3\cos\alpha-1)(1-\cos\alpha)}{(3\cos\alpha+1)(1+\cos\alpha)} .
\eqno(6.9)
$$
For example, for $\alpha=\pi /6$, we obtain
$$
R=\frac{(3\sqrt2-2)(2-\sqrt3)}{(3\sqrt2+2)(2+\sqrt3)}=0.0319 .
\eqno(6.10)
$$
The value is in good agreement with the observed value (6.8). 
By putting $ m_{\nu 3} = \sqrt{ \Delta m^2_{atm} }$, we obtain
$m_{\nu 1}= (0.38\pm 0.02) \times 10^{-2}$ eV, 
$m_{\nu 2}= (0.99^{+0.08}_{-0.05}) \times 10^{-2}$ eV, and 
$m_{\nu 3}= (5.23^{+0.40}_{-0.25}) \times 10^{-2}$ eV.

However, the theoretical reason for $\alpha=\pi /6$ is unclear. 
Since we have assumed the universality of the coupling constants
in the interactions (4.1) and (5.2), rather, 
the case with the universal coupling $y_0^R=y_1^R$,
$$
H_R=y_R \left( \frac{\overline{\nu}_\sigma \nu_\sigma^\ast+\overline{\nu}_\pi 
\nu_\pi^\ast+\overline{\nu}_\eta \nu_\eta^\ast}{\sqrt3}\Phi_\sigma
+\frac{\overline{\nu}_\eta \nu_\eta^\ast-\overline{\nu}_\pi \nu_\pi^\ast}
{\sqrt2}\Phi_\eta
-\frac{\overline{\nu}_\pi \nu_\eta^\ast+\overline{\nu}_\eta \nu_\pi^\ast}
{\sqrt2}\Phi_\pi \right) ,
\eqno(6.11)
$$
is likely. The case corresponds to $\cos\alpha=\sqrt{2 /3}$, and it predicts
$$
R=\frac{4 \sqrt6-9}{ 4\sqrt6+9} =0.0424 .
\eqno(6.12)
$$
The value (6.12) is somewhat large compared with the observed value (6.8), 
but, at present, the case cannot be ruled out within three sigma.
Again, by putting $ m_{\nu 3} = \sqrt{ \Delta m^2_{atm} }$, we obtain
$$
m_{\nu 1}= (0.53^{+0.04}_{-0.03}) \times 10^{-2} {\rm eV} , \ \ 
m_{\nu 2}= (1.17^{+0.08}_{-0.05}) \times 10^{-2} {\rm eV} , \ \ 
m_{\nu 3}= (5.23^{+0.40}_{-0.25}) \times 10^{-2} {\rm eV} . 
\eqno(6.13)
$$

Anyhow, the predicted value of $m_{\nu 1}$ in the present A$_4$ model
is relatively large compared with that in the S$_3$ model \cite{Koide0605}.
We would like to expect the detection from future double beta experiments.

\vspace{3mm}

{\large\bf 7  Superpotential for 3 flavor scalars}

So far, we have not considered the supersymmetric version of the
present model.
Recently, Ma has proposed a SUSY version \cite{Ma0612} of 
the Higgs potential (1.5) which can lead to the VEV relation
$v_\pi^2 + v_\eta^2 =v_\sigma^2$, (1.7).
In a similar way, we can write the superpotential $W$ for the superfields
$\phi^d =(\phi^d_{\bf 1}, \phi^d_{{\bf 1}'}, \phi^d_{{\bf 1}''})$
(hereafter, for convenience, we will drop the index $d$) 
by assuming as follows:

\noindent
(i) The field $\phi_a$ ($a={\bf 1}, {\bf 1}', {\bf 1}''$) to the power
$n$th, $(\phi_a)^n$ ($n=1,2,3$), appears always accompanied  with 
the factor $1/n!$ in the superpotential $W$.

\noindent
(ii) In order to forbid unwelcome A$_4$ invariant terms,
we require
that $W$ is invariant under the transformation
$$
\phi_{{\bf 1}'} \rightarrow -\phi_{{\bf 1}'}, \ \ 
\phi_{{\bf 1}''} \rightarrow -\phi_{{\bf 1}''}. 
\eqno(7.1)
$$
Under this requirement, the terms  $(\phi_{{\bf 1}'})^3$ and 
$(\phi_{{\bf 1}''})^3$ are forbidden, but 
$\phi_{{\bf 1}'}\phi_{{\bf 1}''}$, $(\phi_{{\bf 1}'})^2$ and 
$(\phi_{{\bf 1}''})^2$ are not forbidden.

\noindent
(iii) The A$_4$ symmetry is softly broken by a term $W_{SB}$.

\noindent
As a result, we obtain the superpotential 
$$
W= m\left( \phi_{{\bf 1}'} \phi_{{\bf 1}''} +\frac{1}{2!} \phi_{{\bf 1}}^2
\right) + \lambda \left( \phi_{{\bf 1}}\phi_{{\bf 1}'} \phi_{{\bf 1}''}
+\frac{1}{3!} \phi_{{\bf 1}}^3 \right) +W_{SB} ,
\eqno(7.2)
$$
$$
W_{SB} = \varepsilon m \left( - \phi_{{\bf 1}'}\phi_{{\bf 1}''} 
+\frac{1}{2!} e^{i\theta} \phi_{{\bf 1}'}^2
 +\frac{1}{2!} e^{-i\theta} \phi_{{\bf 1}''}^2 \right).
\eqno(7.3)
$$
Here, although the first and second terms in $W_{SB}$ do not break the A$_4$
symmetry, we have added those to $W_{SB}$ in order to keep
the result (1.7) independent of $W_{SB}$.
We will show below that the superpotential (7.2) can lead to the VEV
relation (1.7) independently of $W_{SB}$ and the parameter $\theta$ in
$W_{SB}$ determines the ratio $v_\pi/v_\eta$, so that the charged lepton
mass spectrum is completely determined only by the parameter $\theta$.

The superpotential (7.2) can be rewritten in terms of the superfields
$(\phi_\pi, \phi_\eta, \phi_\sigma)$ defined by Eqs.(2.9)--(2.11) as follows:
$$
W = m(1-\varepsilon) (\phi_\eta^2 +\phi_\pi^2) 
+\frac{1}{2} m  \phi_\sigma^2 
+\varepsilon m e^{i\theta} (\phi_\eta^2 -\phi_\pi^2 -2 i \phi_\eta \phi_\pi)
$$
$$
+\varepsilon m e^{-i\theta} (\phi_\eta^2 -\phi_\pi^2 +2 i \phi_\eta \phi_\pi)
+\frac{1}{2} \lambda \phi_\sigma \left( \phi_\eta^2 +\phi_\pi^2
+\frac{1}{3} \phi_\sigma^2 \right) .
\eqno(7.4)
$$
Since
$$
\frac{\partial W}{\partial \phi_\pi} = \left[ m(1-\varepsilon)
+\lambda \phi_\sigma - 2 \varepsilon m \cos\theta \right] \phi_\pi
+ 2\varepsilon m\sin\theta \phi_\eta ,
\eqno(7.5)
$$
$$
\frac{\partial W}{\partial \phi_\eta} = \left[ m(1-\varepsilon)
+\lambda \phi_\sigma + 2 \varepsilon m \cos\theta \right] \phi_\eta
+ 2\varepsilon m\sin\theta \phi_\pi ,
\eqno(7.6)
$$
$$
\frac{\partial W}{\partial \phi_\sigma} = m \phi_\sigma
+\frac{1}{2} \lambda (\phi_\eta^2 +\phi_\pi^2 +\phi_\sigma^2) ,
\eqno(7.7)
$$
the minimization conditions of the potential leads to the relations
$$
\tan\theta =\frac{2 v_\eta v_\pi}{v_\eta^2 -v_\pi^2} ,
\eqno(7.8)
$$
$$
v_\pi^2 +v_\eta^2 = v_\sigma^2 ,
\eqno(7.9)
$$
$$
m +\lambda v_\sigma =0 .
\eqno(7.10)
$$
Note that the derivation of the relation (7.8) is independent of
the explicit values of $m$, $\lambda$ and $\varepsilon$, and
the derivation of the relation (7.9) is independent of
the explicit values of $m$, $\lambda$, $\varepsilon$ and $\theta$.
[Also note that the conditions (7.5)--(7.7) can lead to an alternative
solution with $(1-2\varepsilon)(v_\pi^2 +v_\eta^2)=(1+2\varepsilon)
v_\sigma^2$ and $(1-2\varepsilon)m +\lambda \phi_\sigma=0$ instead of
Eqs.(7.9) and (7.10), respectively.  However, we have taken only
the solution which is independent of the parameter $\varepsilon$.]

From the observed values  \cite{PDG06} of the charged lepton masses, 
we obtain
the numerical values $z_1 =0.016473$, $z_2=0.236869$ and
$z_3=0.971402$, where the parameters $z_i$ are defined by
$\sqrt{m_{ei}}=z_i v_d$ with $z_1^2+z_2^2+z_3^2=1$, so that, for
the VEVs of $\phi_a$ defined by Eq.(1.6), we obtain
$z_\pi=0.519393$, $z_\eta=-0.479824$ and $z_\sigma=0.707106$.
Therefore, we can obtain the value of $\theta$ as follows:
$$
\tan\frac{\theta}{2} = \frac{v_\pi}{v_\eta} =
\sqrt{3} \frac{z_3-z_2}{2z_1 -z_2 -z_3} = -1.082466,
\eqno(7.11)
$$
i.e. $\theta=-94.5354^\circ$.
When we express the parameter $z_i$ as
$$
\begin{array}{lll}
z_{1}=\frac{1}{\sqrt6}-\frac{1}{\sqrt3}\sin\xi_e , \\
z_{2}=\frac{1}{\sqrt6}-\frac{1}{\sqrt3}\sin(\xi_e+\frac{2}{3}\pi) , \\
z_{3}=\frac{1}{\sqrt6}-\frac{1}{\sqrt3}\sin(\xi_e+\frac{4}{3}\pi) ,
\end{array}
\eqno(7.12)
$$
the angle $\theta$ is related to the parameter $\xi_e$ by
$$
\frac{\theta}{2} = \xi_e -\frac{\pi}{2}.
\eqno(7.13)
$$
Note that the model gives $m_e \rightarrow 0$ in the limit of
$\theta \rightarrow -\pi/2$. 

\vspace{3mm}

{\large\bf 8 \ Summary}

In conclusion, on the basis of the A$_4$ symmetry, we have investigated a 
Frogatt-Nielsen type model (3.1). The Higgs potential (2.16) for the scalars 
$\phi^d_i$ which are assigned to $(\phi_{\bf 1}, \phi_{{\bf 1}'}, 
\phi_{{\bf 1}''})$ of A$_4$ can lead to the VEV relation (1.7), 
$v_\pi^2+v_\eta^2=v_\sigma^2$, i.e. to the relation (1.4) for 
the VEVs $\langle \phi_i^d \rangle$ defined in (1.6). 
Since the charged lepton 
interactions $\overline{\ell}_L H_L^d \phi^d \phi^d e_R$ give $m_{ei}\propto 
\langle \phi_i^d \rangle^2$, we have obtained the charged lepton mass relation 
(1.2). 
For the neutrino sector, we have obtained the tribimaximal mixing (1.1)
by assuming $ {\bf 3}$ of A$_4$ for the scalars $\phi^u $.

However, it should be noted that, in order to obtain the above results, 
we have needed to assume the following requirements
in addition to the A$_4$ symmetry: 
(i) the ${\bf 1}' \leftrightarrow {\bf 1}''$ symmetry and 
(ii) the universality of the coupling constants.
Those assumptions are phenomenological ones at present.
On the other hand, recently, Ma \cite{Ma0612} has also proposed a model 
which can lead not only to the tribimaximal mixing (1.1), but also to
the charged lepton mass relation (1.2) by assuming a symmetry $\Sigma(81)$.
In Ma's $\Sigma(81)$ model, such an additional assumption except for 
the symmetry $\Sigma(81)$ has not been required. 
However, in his model, we need somewhat unfamiliar and complicated symmetry 
$\Sigma(81)$.
In contrast with the Ma model, in the present model, we have adopted 
a familiar symmetry A$_4$, and, instead, we have put some intuitive 
assumptions (i) and (ii).
Such an approach with phenomenological assumptions, at present, seems to 
be still useful for future extension of the model rather than a 
rigid theoretical model.

In Secs.5 and 6, a possible neutrino mass spectrum has been discussed 
by assuming $\langle \phi_1^u \rangle=\langle \phi_2^u \rangle
=\langle \phi_3^u \rangle$, where $\phi_i^u$ belong 
to {\bf 3} of A$_4$. 
By assuming the structure (6.3) of the right-handed 
neutrino Yukawa interaction, we can speculate the neutrino mass spectrum 
(6.7). 
The case $\alpha=\pi /6$ is interesting from the phenomenological point 
of view, because the case predicts the ratio 
$\Delta m^2_{solar} /\Delta m^2_{atm}=0.0319$. 
However, the numerical predictions in Sec.6 are not conclusive because
the model has needed some speculative assumptions.

In Sec.7, a SUSY version for the Higgs potential of $\phi^d$ has been
proposed. 
The essential idea is indebted to the Ma model based on a $\Sigma(81)$ 
symmetry \cite{Ma0612}.

The present model will give a suggestive hint on seeking for 
a more plausible model 
which leads to the tribimaximal mixing (1.1) and the charged lepton
mass relation (1.2).

\vspace{4mm}

\centerline{\large\bf Acknowledgements} 

The author would like to thank T.~Singai for helpful
conversations on the A$_4$ symmetry.
He also thank E.~Ma for informing his recent paper \cite{Ma0612}
prior to putting it on the hep-ph arXive.
The author is much indebted to his recent paper \cite{Ma0612}, 
especially, for Sec.7.
This work is supported by the Grant-in-Aid for
Scientific Research, Ministry of Education, Science and 
Culture, Japan (No.18540284).



\newpage

{\bf Table 1} \ A$_4$ assignments of the fields

\vspace{2mm}
\begin{tabular}{|ccc|} \hline
Fields &  A$_4$    &   U(1)   \\ \hline
$\ell_L$ &  {\bf 3}  &  0 \\
$\nu_R$   &  $({\bf 1},{\bf 1}',{\bf 1}'')$ & 0 \\
$e_R$    &  {\bf 3}  &  0 \\ \hline
$\phi^u$ & {\bf 3}   &  $q_u$  \\
$\phi^d$ & $({\bf 1},{\bf 1}',{\bf 1}'')$ & $q_d$ \\
$\Phi$   & $({\bf 1},{\bf 1}',{\bf 1}'')$ & 0 \\
$H_L^u$  & {\bf 1}   & $ -q_u$ \\
$H_L^d$  & {\bf 1}   & $-2q_d$ \\ \hline
\end{tabular}
\end{document}